\documentclass{aa} %
\usepackage{graphicx}
\usepackage{txfonts}
\usepackage{natbib}
\begin{document}
   \title{Estimating the redshift of PKS~0447$-$439 through its GeV--TeV emission}

   \author{E.~Prandini
          \inst{1},
          G.~Bonnoli\inst{2},
          and
          F.~Tavecchio\inst{2}
          }

   \institute{Dipartimento di Fisica, Universit\`{a} di Padova, Via Marzolo 8, I-35131, Padova, Italy \\
         \and
             INAF – Osservatorio Astronomico di Brera, via E. Bianchi 46, I-23807 Merate (LC), Italy \\
             }

   \date{Received -- ; accepted --}
  
 
  \abstract
   {Blazars are radio-loud active galactic nuclei (AGN) with a jet pointing at 
     small angles towards the observer. 
     The  overall emitted spectrum is typically non-thermal,
     and in some cases the emission and/or absorption lines
     are so faint as to prevent the determination of the redshift based on optical spectroscopy methods. 
     PKS~0447$-$439 is a bright blazar recently detected at very high energy. 
     The redshift of the source is still disputed: a recent spectral analysis reports only a 
     lower limit of $z$\,$>$\,1.246, which contradicts
      the previous measure of $z$\,=\,0.205 reported in the literature.
   }
   {We aim to give a redshift estimate of the blazar PKS~0447$-$439 based on combined 
     GeV ({\it Fermi}/LAT) and TeV (H.E.S.S.) observations.}
   {Taking into account the absorption of TeV photons by 
     the interaction with the extragalactic background light (EBL),
     we applied the method we developed in a previous work to derive the 
     redshift of PKS~0447$-$439.
     Moreover, we compiled the overall spectral energy distribution (SED)
     using  optical-UV, soft X--ray, and 
     $\gamma$-ray data, nearly simultaneous to the H.E.S.S. observations at TeV energies.
     Finally we modelled the spectral energy distribution (SED) within the framework of a homogeneous, 
     leptonic synchrotron self-Compton (SSC) model.
   }
   {Using the recent TeV spectrum measured by H.E.S.S. 
    we obtain for PKS~0447$-$439 a redshift of
    $z_{rec}$\,=\,0.20\,$\pm$\,0.05, which
    is our estimate on the source distance. 
    This value agrees very well with the value 
    reported in the literature  and confirms that our method can be
    successfully used to constrain blazars distances.
    Assuming this distance, the SED can be well fitted with the above mentioned model. 
    The physical parameters that we find
    suggest a strongly matter-dominated jet.}
  {
    Our analysis confirms that the redshift of PKS~0447$-$439 is likely 0.2, and 
    supports the result present in the literature.
  }
  
 
  \authorrunning{E. Prandini et al.}
  \maketitle
  %
  
  \section{Introduction}
  
  The number of known AGN whose emission extends
  up to the very high energy (VHE; E~$>$~100\,GeV) 
  $\gamma$--ray band has more than doubled in the last
  couple of years. Nowadays, the TeV sources catalogue counts 44  
  AGN\footnote{See TeVCat (http://tevcat.uchicago.edu/) and R.~Wagner (http://www.mpp.mpg.de/$\sim$rwagner/sources/) pages.}, located both in 
  the northern and in the southern hemisphere.
  This achievement has been made possible thanks to the
  good performances of the last generation of imaging atmospheric Cherenkov
  telescopes (IACT), namely MAGIC,
  H.E.S.S. and VERITAS.
  A key ingredient for the detection of new sources has been
  the cooperation of these telescopes with satellite experiments, 
  especially those operating at optical, X--ray and soft $\gamma$--ray frequencies.
  
  The great majority of AGN detected at VHE belongs to the class of blazars. In these
  sources, the emission is produced in relativistic jets that point towards the
  observer.
  
  PKS~0447$-$439 is a blazar located at $\rm{RA}(J2000)\,=\,04^{\rm{h}} 49^{\rm{m}} 24\fs88$,~
  $\rm{Dec.}(J2000)\,=\,-43\degr 50\arcmin 09\farcs 7$.
  It was discovered in 1981 at radio wavelenghts by the Molonglo Telescope 
  \citep{large81} and detected by the PMN radio Survey in 1993 \citep{gregory94}.  
 
  The source was subsequently detected  in the UV with EUVE \citep{lampton97} and X--rays with ROSAT \citep{white94}.
  Some confusion arose in the determination of the  optical counterpart because
  at
  first a near--UV bright spectrum with prominent emission
  lines in the optical was related to PKS0447-439 \citep{craig97}. 
  The source was therefore  classified as a Type~I 
  Seyfert at $z$\,=\,0.107. Instead \citet{perlman98} reported a featureless
  optical continuum typical of a BL Lac object.
  
  We investigated this classification discrepancy, 
  also inspecting some archival  {\it Swift}/UVOT exposures of the field. We understood 
  that the association of the Seyfert spectrum with this source is probably due to some 
  databasing mistake, because the PKS~0447$-$439 coordinates reported in \citet{craig97} are
  incorrect, and point to a region that is source--free both in their optical
  finding chart and in our UVOT images.

  The redshift of PKS~0447$-$439 was measured by 
  \citet{perlman98}, who reported a value of $z\,\sim\,$0.205 based on few weak
  absorption features in the optical spectrum that were
  interpreted as the $Ca$H,K doublet.
  Another study performed by \citet{landt08} provided only a lower limit
  of 0.176, based on the non--detection of the host galaxy, which is 
  assumed to be a
  giant elliptical of inferred luminosity \citep{piranomonte07}. 
  Recently, a different lower limit z $>$ 1.246  was obtained \citep{landt12} based on the analysis of absorption lines in the optical spectrum. This new value is well above the measurement reported by  \citet{perlman98}.

  The Large Area Telescope (LAT) instrument on board {\it Fermi} 
  identified the source as one of the brightest of the 
  southern hemisphere in the 100\,MeV--300\,GeV energy range \citep{abdo09}.
  Dedicated studies on the weekly light curve above $300\,$MeV 
  revealed that the source is variable in this energy range \citep{abdo10a}.  
  Inspired by the {\it Fermi}/LAT observations, the H.E.S.S. IACT observatory
   performed a dense observation campaign between November 2009 and January 2010.
  This led to the detection of a VHE signal, statistically significant above
  the 13\,$\sigma$ level and positionally consistent with PKS~0447$-$439, as 
  preliminarily reported in \citet{zech11}. 
  
  In this Paper, we derive an independent estimate of the 
  distance to PKS~0447$-$439 using combined GeV and TeV spectral 
  information, following the method we proposed in \citet{prandini10}. 
   Our goal is  to provide a new measurement of the source's 
  redshift, which is still uncertain, as discussed above.
  In the next section we briefly outline the method and present the
  result, which basically confirms the estimate of  $z\,\sim\,$0.2 
  given by \citet{perlman98}. 
  Then, we  build a quasi-simultaneous SED, which is fitted with a
  homogeneous, leptonic  SSC model. This allows the determination of 
  the main physical parameters governing the jet physics.

  \section{Inferring the distance of TeV emitting blazars}
  The VHE  spectrum of blazars suffers from the
  absorption arising from the interaction with the extragalactic background light (EBL)
 \citep{hauser01}.
  This absorption, caused by electron-positron pair
  production \citep{nikishov62}, 
  induces a partial/total deformation of the VHE part of the spectrum that is
  strongly redshift-dependent. In general, for nearby sources 
  that are located at redshift below 0.1, it affects 
  the spectral points above some TeV. At higher
  redshifts, between 0.1 to 0.5, it 
  affects the spectrum already at some hundred of GeV, while above 0.5  
  it becomes effective already above some GeV. 
  Therefore, given an intrinsic VHE
  spectrum, the observed one depends on the distance of the emitter.
  In other words, the optical depth associated to the absorption process
  depends on the distance covered by the energetic photon 
  and on its energy.
  Unfortunately, owing to strong foreground emissions that are 
  difficult to suppress,
  there are no solid measurements of EBL. Upper and lower limits are provided by
  indirect techniques such as galaxy counts, which provide solid lower
  limits.
  In addition to these measurements, many models of the EBL energy density 
  and evolution have been proposed in the last years 
  \citep{stecker06,franceschini08,kneiske10,dominguez11}. 
  In this work we adopt the model presented in \citet{franceschini08}.

  To estimate the distance to PKS~0447$-$439, 
  we applied the method that we proposed in  \citet{prandini10} 
  and updated in \citet{prandini11}. This method is based on the comparison
  between the high energy (HE; $0.1 < E < 100\,$ GeV) $\gamma$--ray spectrum measured 
  by {\it Fermi}/LAT and  that measured at VHE by the 
  last generation of imaging atmospheric Cherenkov telescopes (MAGIC,
  H.E.S.S. and VERITAS). 

  We refer the reader to the original paper \citep{prandini10}
  for the full description of the method.
  Briefly, to infer the distance of an HE and VHE gamma-ray emitting
  blazar, we calculate the redshift, $z^*$, at which 
  the power law slope fitting the VHE spectrum corrected for the
  EBL absorption equals the slope measured by {\it Fermi}/LAT at lower energies. 
  As empirically demonstrated in the paper using a set of known redshift sources,
  this value is related to  the true redshift of the source by a simple linear relation.
  Therefore, once one has obtained $z^*$,  it is possible to give an estimate on the source distance, 
  $z_{rec}$, by using the inverse formula:
  \begin{equation}\label{eq:1}
    z_{rec} =  \frac{(A - z^*)}{B}, 
  \end{equation}
  with A and B given in \citet{prandini11}. 
  The law was applied to infer the distance of the unknown redshift blazar PKS~1424$+$240.
  The value obtained has been confirmed in a study on the host galaxy 
  performed by \citet{meisner10}.

  \subsection{The reconstructed $z$ of PKS~0447$-$439}
\begin{figure}
    \centering
    \includegraphics[width=3.5in]{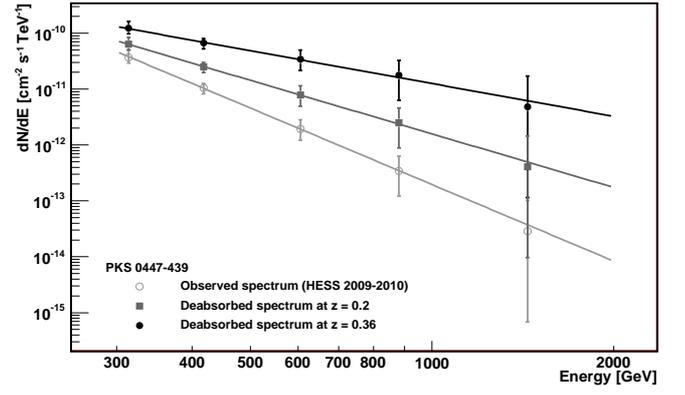}
    \caption{Observed (light grey, open circles) and deabsorbed VHE spectra of PKS~0447$-$439, as reported
      in \citet{zech11}. Deabsorption is applied assuming the \citet{franceschini08} EBL model, and a redshift of the source $z^*=0.36$ (black, filled circles) and $z_{rec}=0.2$ (dark grey, filled squares).} 
    \label{FigGam}
  \end{figure}

  The blazar PKS~0447$-$439 was observed by the H.E.S.S. IACT observatory
  between November 2009 to January 2010, for a total of 13.5 hours 
  of data.
  A clear signal of 13.8\,$\sigma$ of significance was detected at energies 
  above 250\,GeV \citep{zech11}. The preliminary spectrum extends from
  300\,GeV up to more than 1\,TeV, and is compatible with a steep simple power law of index 
  $\Gamma$\,=\,4.36\,$\pm$\,0.49, where the error reported is statistical only. 
  The significant spectral points are 
  drawn in Fig.~\ref{FigGam}, open markers.
  
  The {\it Fermi}/LAT spectral slope in the energy range 0.1-100\,GeV
  is $\Gamma_{LAT}$\,=\,1.95\,$\pm$\,0.03, as reported in the 1FGL catalogue
  \citep{abdo10b}.
  We adopted this slope, and not the one extracted from only simultaneous data, 
  to be 
  consistent with the analysis performed for the determination of
  the $z^*\,-\,z_{true}$ law. 
  
  The redshift that we obtain when requiring that the deabsorbed 
  PKS~0447$-$439 spectrum has a slope equal to $\Gamma_{LAT}$ is
  $z^*$\,=\,0.357\,$\pm$\,0.065, where the error takes into account both the errors 
  on the TeV slope and on the {\it Fermi}/LAT slope.
  The corresponding spectrum is drawn in Figure~\ref{FigGam}, 
  filled black circles. 
  A simple power law  fits the data very well, 
  with a probability higher than 99$\%$, and $\chi^2/$ndf\,=\,0.1/3.
  
  \begin{figure}
    \centering
    \includegraphics[width=3.5in]{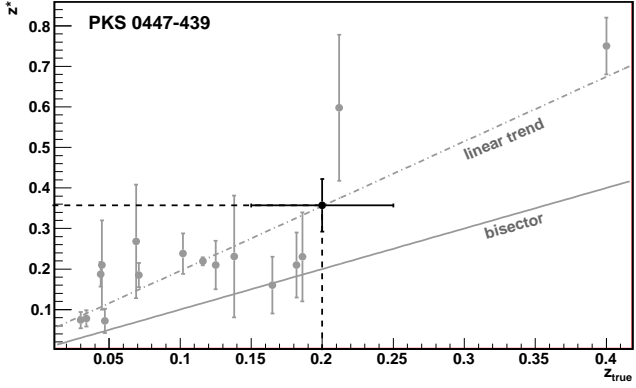}
    \caption{Diagram adapted from \citet{prandini11} representing the linear relation between 
      the redshifts of known distance sources and their \emph{z$^*$} values. The value of 
      \emph{z$^*$} obtained for PKS~0447$-$439 in this work is superimposed in black.}
    \label{LinearRelation}
  \end{figure}
  
  Figure~\ref{LinearRelation} shows the  $z^*\,-\,z_{true}$ plot in
  linear scale, adapted from \citet{prandini11}. The grey markers represent the
  $z^*$ values of the sixteen sources with known redshifts
  used to infer the linear relation (dashed-dotted line).
  The continuous line is the  $z^*\,=\,z_{true}$ line. 
  All points lie above or on the line, 
  meaning that the values $z^*$  are higher than the real redshifts. This 
  indicates $z^*$ as  a good upper limit on the source distance.
  In our case, we have $z^*$ of 0.357\,$\pm$~0.065 and the corresponding upper limit is 0.49
  at 2 sigma level.
  Applying Equation~\ref{eq:1} to our data, we obtain for 
  PKS~0447$-$439 the distance $z_{rec}$\,=\,0.20\,$\pm$\,0.05, where the error
  is statistical only.
  The differential energy spectrum obtained assuming a distance 
  $z$ of 0.20 is drawn in Figure~\ref{FigGam}, filled squares. 
  It lies between the observed spectrum (not corrected for EBL absorption)
  and the spectrum obtained assuming a distance $z^*$. A simple power 
  law of index $\Gamma$\,=\,3.2\,$\pm$\,0.5 fits the data very well
  (probability $>$~99$\%$).

\begin{figure*}
    \centering
    \includegraphics[width=0.85\textwidth]{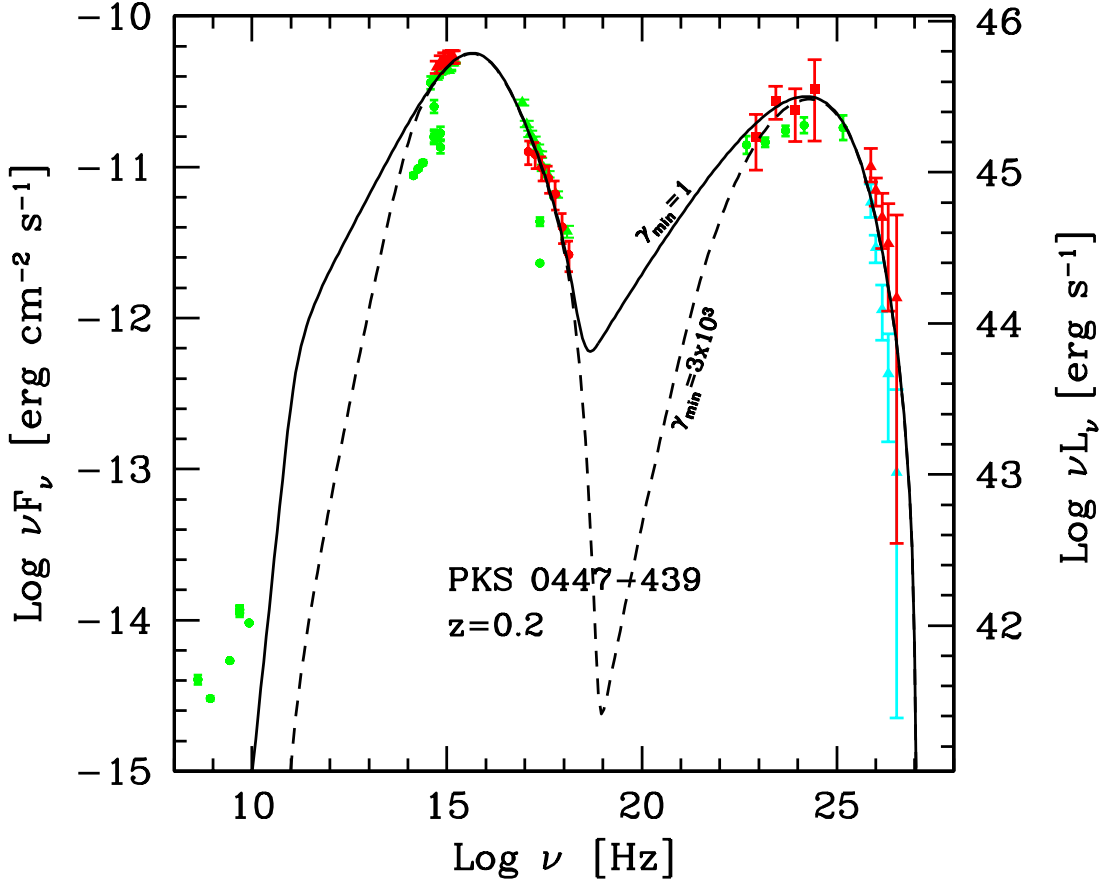}
    \vspace{-2.0cm}
    \caption{SED of PKS~0447$-$439 during the epoch of H.E.S.S. observation 
      (November 2009 $-$ January 2010). Red points represent {\it Swift}/UVOT, 
      {\it Swift}/XRT, {\it Fermi}/LAT and de-absorbed (assuming $z\,=\,0.2$) 
      H.E.S.S. data. Cyan data are the observed H.E.S.S. data. Green symbols 
      report historical data obtained through the ASI/ASDC tools. Green LAT 
      points are from the 1LAC catalogue. The black lines are the result of the 
      one-zone emission leptonic model discussed in the text for two different 
      values of the minimum Lorentz factor of the emitting electrons,  
      $\gamma _{\rm min}\,=\,1$ (solid) and $\gamma _{\rm min}\,=\,3\times 10^3$ (dashed). See text for details.}
    \label{fig:sed_0447}
  \end{figure*}
  
  The estimated distance obtained perfectly agrees
  with the value $z$\,=\,0.205 reported by \citet{perlman98}, 
  and also confirms the lower limit of $z~>$~0.176 based on 
  photometric estimates.
  Therefore, our study strongly supports the value of $z$\,=\,0.205
  for PKS~0447$-$439.

 \begin{table*}[th]
    \centering
    \begin{tabular}{lcccccccccccc}
      \hline
      \hline
      $\gamma _{\rm min}$ & $\gamma _{\rm b}$ & $\gamma _{\rm max}$ & $n_1$ & $n_2$ &$B$ & $K$ &$R$ & $\delta $ &$P_{\rm e}$ & $P_{B}$ & $P_{\rm p}$ & $P_{\rm r}$ \\
       & [$ 10^4$] &[$ 10^5$]  &  & &[G] & [$ 10^3$ cm$^{-3}]$  & $[10^{16}$ cm] & &  [$10^{45}$ erg/s] & [$10^{45}$ erg/s]& [$10^{45}$ erg/s] & [$10^{43}$ erg/s]\\
      \hline
      $1$ & $2.6$ & $3$ & $2.0$ & $4.4$ & $0.07$ & $5$ & $1.45$ & $39$  & 1.7 & $7.8\times 10^{-3}$& 230 & 3.8\\
      $10^3$ & $2.6$ & $3$ & $2.0$ & $4.4$ & $0.07$ & $5$ & $1.45$ & $39$  & 0.52 & $7.8\times 10^{-3}$& 0.07 & 3.6\\

      \hline
      \hline
    \end{tabular}
    \vskip 0.4 true cm
    \caption{Input model parameters for the models reported in Fig.~3. 
      We report the minimum, break and maximum Lorentz factor
      and the low- and high-energy slope of the electron energy distribution, the magnetic field intensity, the
      electron density, the radius of the emitting region and its Doppler factor. We also report the derived power
      carried by electrons, magnetic field, protons (assuming one cold proton per emitting relativistic electron) and the total radiative luminosity.}
    \label{param}
  \end{table*}

   \section{Spectral energy distribution}
  The single-epoch multiwavelength SED of PKS 0447$-$439 is plotted in
  Fig.~\ref{fig:sed_0447}, and includes, in addition to the H.E.S.S. 
  data presented in previous section,
  {\it Swift}/UVOT, {\it Swift}/XRT, and {\it Fermi}/LAT data acquired during 
  the epoch of the H.E.S.S. detection, and other historical data.
 
  {\it Swift} observed PKS~0447$-$439 for about one week at the end of the H.E.S.S.
  campaign \citep{zech11}.  
  Data from the UVOT \citep{roming05} observation taken on December 25
  (obs. ID 00038100009) were analysed 
  by means of  the \texttt{uvotimsum} and \texttt{uvotsource} 
  tasks with a source region of $5''$, while the background was extracted from a
  concentric source--free annular region with inner and outer radii of $10''$
  and $15''$, respectively. 
  The extracted  magnitudes were corrected for Galactic
  extinction  using the values of \citet{schlegel98} and applying the
  formulae by \citet{pei92} for the UV filters, and eventually
  were converted into fluxes following \citet{poole08}. 
  
  XRT \citep{burrows05} $\nu F_{\nu}$ points corresponding to the observation
  of Dec 25 (MJD 55190, obs. ID 00038100010) were obtained through the
  ASI/ASDC on-line analysis tool\footnote{http://tools.asdc.asi.it/}
  \citep{stratta10}. The X--ray fluxes were obtained assuming a power law
  spectrum (best-fit photon index $\Gamma\,=\,-2.46$) and fixing the absorption to
  the Galactic value $N_{\rm H}^{Gal}\,=\,1.24 \times 10^{20}$ cm$^{-2}$
  \citep[after][]{kalberla05}.
  
  We retrieved  the publicly available data
  taken by the LAT $\gamma$-ray telescope on board  the \emph{Fermi} satellite
  \citep{atwood09} in scanning mode from the NASA database\footnote{http://fermi.gsfc.nasa.gov/}. 
  We selected the good quality (``DIFFUSE'' class) events observed within
  10$^{\circ}$ from the source position, taken between MJD 55170 and MJD
  55190,
  and with measured energy in the  0.2--100\,GeV interval. We excluded events
  observed
  at zenith distances greater than 105$^{\circ}$ to avoid
  contamination from  the Earth albedo.
  We performed the analysis by means of the standard science tools, v. 9.18.6,
  including Galactic and isotropic extragalactic backgrounds and the P6 V3
  DIFFUSE instrumental response function.
  We applied  an unbinned likelihood algorithm
  (\texttt{gtlike}) to the data, 
  modelling the source spectrum with a power law model,
  with the integral flux in the 0.2--100\,GeV energy band  and photon index left as
  free parameters. We clearly detected the source with a test statistics
  \citep{mattox96} ${\rm TS}\,=\,154$, and determined $F_{0.2-100 {\rm
      GeV}}\,=\,6.93\,\pm\,1.26 \times 10^{-8}$  ph\,cm$^{-2}$s$^{-1}$
  and $\Gamma\,=\,1.93\,\pm\,0.14$.  
 
  We also performed the same analysis on the dataset split into logarithmically
  equal bins of energy, accepting spectral points with TS $> 25$ 
  and more than three
  events attributed to the source in the model. The results are plotted in
  Figure~\ref{fig:sed_0447}.

  PKS 0447-439 belongs to the group of highly peaked BL Lac objects. 
  Within the framework of leptonic models, the emission of this type of 
  sources is generally modelled with the one-zone
  SSC model \citep[e.g.][]{tavecchio98,tavecchio10,abdo11}. 
  The idea that the emission mainly originates in a single, uniform, 
  region is supported by the correlated variability observed at different 
  frequencies, especially X-rays and gamma-rays \citep[e.g.][]{fossati98}.

  The good spectral coverage allows one to firmly constrain the 
  low-energy bump, 
  which is related to synchrotron radiation produced
  by accelerated electrons in the jet,  
  whose peak lies around $10^{16}$ Hz (see Fig.~3). 
  LAT and H.E.S.S. data track the so-called SSC component quite well, 
  which shows a broad peak in the GeV region. 
  As detailed in \citet{tavecchio98}, in the framework of the one-zone 
  SSC model knowing of the peak frequencies and luminosities, 
  together with an estimate of the variability timescale, allows one to 
  determine the physical  parameters of the emitting region. 
  An indication of the variability timescale can be derived 
  from the multi-frequency observations, especially those at X-ray energies, 
  which reveal variability 
  on timescale of $\approx$\,days. We therefore assume an upper limit of $t_{\rm var}<1$ day for the variability timescale.
  
  We modelled the SED with the one-zone leptonic model described in
  \citet{maraschi03}. The emission region is spherical with radius $R$, in
  motion with bulk Lorentz factor $\Gamma$ at an angle $\theta $ with respect to
  the line of sight. Special relativistic effects are fully accounted for by the
  relativistic Doppler factor, $\delta\,=\,[\Gamma(1-\beta \cos \theta)]^{-1}$. The
  energy distribution of the emitting electrons is assumed to be well described
  by a smoothly connected broken power law function,  with minimum, maximum and break Lorentz 
  factor  $\gamma _{\rm min}$, $\gamma_{\rm max}$ and $\gamma _{\rm b}$, respectively. The SSC emission 
  is calculated assuming the full Klein-Nishina cross section \citep{jones68}. 
  We recall that within the framework of the one-zone SSC model the synchrotron 
  self-absorption causes the source to be opaque below frequencies of about
  $10^{11}-10^{12}$ Hz. The emission below these frequencies 
  is therefore produced by more distant, transparent regions of the jet.
  
  Two possible SEDs are reported in Fig.~3, 
  while the corresponding input 
  parameters are listed in Table~\ref{param}. 
  In the table we also report the derived powers carried by the different components, 
  relativistic electrons, magnetic field and protons (assuming a composition of one 
  cold proton per relativistic electron) and the total radiative power of the jet, 
  $P_{\rm r}\simeq L_{\rm obs}/\delta^2$ (in which we assumed $\delta\sim \Gamma$).

  The jet power is strongly dependent on the total density of particles in the jet that, 
  in turn, is dominated by the number of relativistic electrons at $\gamma _{\rm min}$. 
  While this parameter can be relatively well determined for FSRQs, 
  because the 
  low-energy end of the electron energy distribution 
  is directly accessible through the 
  inverse Compton emission that falls in the X-ray band 
  \citep[e.g.][]{ghisellini10},
  for BL~Lac objects 
  emitting through SSC the constraints are quite loose. 
  As an example of the impact of 
  different values of $\gamma _{\rm min}$ on the derived SED, we report in Fig.~\ref{fig:sed_0447}
  two curves, corresponding to $\gamma _{\rm min}\,=\,1$ (solid line) and $\gamma _{\rm min}\,=\,3\times 10^3$ 
  (dashed). The two curves clearly show that the value of $\gamma _{\rm min}$ mainly affects 
  the low-energy branch of the synchrotron and SSC bumps. Clearly, any value between 1 and 
  $10^3$ is allowed by the available data, which determines 
  a large uncertainty on the inferred powers (see Table 1).  
     
  The magnetic field energy density appears to be several orders of magnitude below the equipartition value with 
  the electron energy density and, consequently, the corresponding Poynting flux is negligible compared 
  to the power carried by protons and electrons. This is a feature that PKS~0447$-$439 shares with several 
  other TeV BL Lacs \citep[e.g.][]{tavecchio10,ghisellini10}.

  \section{Conclusions}

  We have presented the results of our study
  on the distance to PKS~0447$-$439.  
  This work was triggered by the preliminary measurement of the differential
  energy spectrum emitted by the source at VHE gamma-rays
  performed by the H.E.S.S. Collaboration \citep{zech11}.
  The redshift we found with our method, based on both HE and VHE spectra,
  perfectly agrees with the measurement 
  performed by \citet{perlman98}, which reports a redshift of 0.205. 
  But it contradicts with the recent assertion
  of \citet{landt12}, who claimed a redshift higher than 1.246.

  Assuming this redshift, we have built a broadband SED including quasi-simultaneous
  ultraviolet, X-ray and gamma-ray data acquired during 
  the epoch of the H.E.S.S. detection.
  The SED can be well fitted within the framework of
  a homogeneous, leptonic SSC model. 
  The available data leave a large uncertainty on one of the 
  parameters of the model, $\gamma _{\rm min}$, the
  minimum Lorentz factor of the emitting electrons. 
  This uncertainty propagates to the inferred powers. 
  In any case, the physical parameters that we found 
  suggest a strongly matter-dominated jet, in agreement with
  the other sources observed at these extreme energies.

  The method used in this paper has great potential.
  A large part of GeV emitting blazars, indeed, about 
  60\% of the BL Lac objects present in the 
  second catalogue of AGN detected by {\it Fermi}/LAT \citep{ackermann12},
  has unknown redshift. 
  If detected in the VHE band, the redshift of these powerful objects
  could be investigated through this technique. 
  Therefore, we encourage the observation of
  promising TeV emitters among the unknown/uncertain redshift 
  {\it Fermi} blazars.

  
  \begin{acknowledgements}
    This research has made use of public {\it Swift} and {\it Fermi} data obtained from the High Energy Astrophysics Science Archive Research Center (HEASARC), provided by NASA’s Goddard Space Flight Center through the Science Support Center (SSC). Part of this work is also based on archival data and on-line services provided by the ASI Science Data Center (ASDC). 
  \end{acknowledgements}

\end{document}